# On the compatibility of the topologies of parallel tasks and computing systems

[1]A.F. Zadorozhny and [2]V.A. Melent'ev

[1]Novosibirsk State University of Architecture and Civil Engineering, 8 Leningradskaya str., 630008, Novosibirsk, Russia
[2]A.V. Rzhanov Institute of Semiconductor Physics, Siberian Branch of the Russian Academy of Sciences, 13 Lavrentyev aven., Novosibirsk, 630090, Russia

*Corresponding author: melva@isp.nsc.ru

**Abstract.** Aspects of compatibility of topologies of parallel computing systems and tasks are investigated. The introduction of appropriate indexes based on the original topological model of parallel computations and on the nontraditional description of a graph by its projections is proposed and elucidated. On the example of hypercubic computing system (CS) and tasks with ring and star information topologies, we demonstrate determining the indexes and their use in a comparative analysis of the applicability of interconnect with a given topology for solving tasks with the same and different types of information topologies.

1. Introduction

Based on Amdahl Law [1], the model allows estimating the limit acceleration of the tasks that have scalarity (nonparallelizability) of separate fragments. However, the abstraction of such a model from the task's information topology and from the interconnect system topology hampers getting a complete picture for the conditionality of the task's parallelism from the system topologies and, hence, the formalized optimization of the architecture for the latter under the classes of solved tasks. The proposed report is devoted to the study of the compatibility of topologies of parallel systems and tasks. The corresponding indicators based on the original topological model of parallel calculations [2-8] and on the unconventional graph description by projections [9-13] are introduced.

2. Basic provisions and the topologies compatibility concept

The real-time implementation of a parallel application on a CS can be successful, if: 1) its interconnect operation speed is sufficient to ensure that the delays made by the exchange processes between parallel branches do not exceed the delays allowed by real-time requirements and 2) the system graph of supplemented by edges, corresponding to the preceding condition, contains a subgraph, which is isomorphic to the information graph of the task, with the number of vertices (parallel branches) corresponding to the same condition.

The studies of system and network topologies are based on graph representations when bijective correspondences are established between the modules of the system and its vertices, as well as between the connection lines and edges of a graph. The main drawback of the traditionally used matrix-list descriptions is that the adjacency/identity descriptions of graph vertices/edges are identical, whereas the routes and cycles usually used for evaluating the quality of structures are multiplace relations on a set of vertices.

In this regard, in the topological model used here, the authors allow non-adjacent (mediated/indirect) connections of information-related processors. This allows increasing the graph degree of the system and increasing the capability of isomorphic embedding of the graphs of solved tasks in it, the ratio between the volumes of *W* computation and information interactions in the parallel (*W*, *Q*)-the task and the used fast network technology operation speed determining the interdependence of the order of the task graph and the maximum distance, allowed in the graph of the system, between the information-related vertices, i.e. the reachability of $\partial$. The topological compatibility of the task with the number of parallel branches of *p* and CS with a number of processors $n > q$ we is understood as the possibility of isomorphic embedding of the task's information graph of order *p* in the system graph transformed according to the specified reachability value $\partial$ so that the vertices in it are adjacent if the distance between them in the original CS graph of order *n* does not exceed $\partial$.

### 3. Investigation of compatibility on graph models

The essential properties of the ring ($W_R$) and star ($W_Z$) tasks, considered here as an example of topologies of tasks, do not require additional explanations of their names. The choice of problems with such information topologies, as well as the choice of a hypercube as the CS topology, is primarily due to the simplicity and visual demonstration of the method proposed in this paper for evaluating the compatibility of topologically different parallel tasks and systems.

Any hypercube $H_s$ has a Hamiltonian cycle passing through each vertex exactly once [14], and although it (hypercube) is a bipancyclic graph (contains cycles of only even length at $n > 3$) [15], problems with the organization of the cycle per unit less or more derived from the model [7] of the odd *p*-value should not occur. Therefore, regardless of the dimensionality *s* of hypercubic CS, the embedding of a task with a ring topology in it can be absolute even at $\partial = 1$. This means that the use of hypercubic CSs does not introduce any limits in the parallelization of tasks with a ring topology, and if they do occur, they are conditioned only by the sufficient operation speed of the network technology used in CSs.

In [16], the following formula of potential $p_Z(H_s)_\partial$, generalized to any dimensions and reachability $\partial \leq s$ of hypercube $(H_s)_\partial$, was obtained:

$$p_Z(H_s)_\partial = \sum_{i=0}^{\partial} \binom{s}{i}.$$

Using this formula, we will build up a table of $\partial$-compatibility ($0 < \partial < 4$) for tasks with the "star-type" topology with the *s*-dimensional ($1 < s < 9$) hypercubic CS:

| *s* | 2 | 3 | 4 | 5 | 6 | 7 | 8 |
|---|---|---|---|---|---|---|---|
| $n = 2^s$ | 4 | 8 | 16 | 32 | 64 | 128 | 256 |
| $p_Z(H_s)_1$; $C_Z(H_s)_1$ | 3; 0,75 | 4; 0,5 | 5; 0,3125 | 6; 0,1875 | 7; 0,1094 | 8; 0,0625 | 9; 0,0352 |
| $p_Z(H_s)_2$; $C_Z(H_s)_2$ | 4; 1 | 7; 0,875 | 11; 0,6875 | 16; 0,5 | 22; 0,3438 | 29; 0,2266 | 37; 0,1445 |
| $p_Z(H_s)_3$; $C_Z(H_s)_3$ | 4; 1 | 8; 1 | 15; 0,9375 | 26; 0,8125 | 42; 0,6563 | 64; 0,5 | 93; 0,3633 |

From the above table, it is not difficult to make sure of the following:
1. An increase in the interconnect operation speed (and the corresponding increase in the maximum length $\partial$ of allowed connections) leads to a significant increase in the potential of parallelism and compatibility of any topologies of the CS and parallel tasks solved on it.
2. When the hypercube size is increased, its compatibility with the tasks of "star-type" topology on the CS drops significantly due to a very small (compared to the increase in the number *n* processors) increase in the potential of parallelism.

Thus, on the example of the hypercubic CS and tasks with ring and the "star-type" information topologies, the determination of compatibility indicators has been demonstrated, the topological aspects of compatibility of such CSs and tasks have been investigated, and the possibility of using the proposed indexes in the analysis of the applicability of the interconnect of other topologies of systems for solving tasks with the given information topology has been quite clearly demonstrated in the present contribution.

**4. Conclusion**

The topological compatibility index is proposed for the numerical systems of parallel tasks solved on them. The index is abstracted from the technical characteristics of interconnect used in the CS and allows assessing the potential parallelization of a particular task, which is determined only topologically, i.e. by the interconnect topology and the information topology of the task.

The results of this work will be useful both in comparing parallel systems and in optimizing the choice of their topologies, corresponding to the set of the tasks to be solved and when analyzing the potential capabilities of a concrete system in the solution of certain parallel tasks.